# Gender Differences in Injury Severity Risk of Single-Vehicle Crashes in Virginia: A Nested Logit Analysis of Heterogeneity

Md Rauful Islam[1], Kaveh Bakhsh Kelarestaghi[2*], Alireza Ermagun[3], and Snehanshu Banerjee[4]

1 The Charles Edward Via, Jr. Department of Civil and Environmental Engineering, National Capital Region, Virginia Tech. 750 Drillfield Dr., Blacksburg, VA 2406,

2 The Charles Edward Via, Jr. Department of Civil and Environmental Engineering, National Capital Region, Virginia Tech. 900 N Glebe Rd, Arlington, VA 22203.

3 Department of Civil and Environmental Engineering, Mississippi State University, 501 Hardy Road, 235 Walker Hall, MS, 39762, US;

4 Department of Transportation and Urban Infrastructure Studies, Morgan State University, 1700 E. Cold Spring Lane, Baltimore, MD 21251, USA,

*Corresponding Author Email: kavehbk@vt.edu

## ABSTRACT

This paper explores gender differences in injury severity risk using a comprehensive crash dataset including the driver, vehicle, environment, and roadway characteristics. For the purpose of this study, only single vehicle crashes that occurred in the Commonwealth of Virginia, collected in a five-year period of 2011-2015, were used. This study contributes to the literature of crash analysis by incorporating an extensive dataset with normalized attributes. The dataset used for model estimation integrated from two different data sources. These data sources include Virginia Traffic Records Electronic Data System (TREDS) to incorporate crash information and Virginia Base Mapping Program to incorporate roadway characteristics. A two-level nested logit model is developed for each gender, in order to relax the Independence of Irrelevant Alternatives (IIA) assumption. Several crash determinants including but not limited to, driver's characteristics (e.g., age), driver's behavior, vehicle condition, weather conditions, roadway geometry and surface conditions were found to have an impact on both male and female drivers. This study is an assessment of influence of different factors on crash severity segmented by gender. A few interesting results include; (i) female fatality risk exceeds male fatality risk when driving under the influence of alcohol, (ii) speeding contributes to lower fatality risk among female drivers compared to their male counterparts, (iii) poor vehicle conditions have no impact on injury risk to female drivers while having an increased injury risk among male drivers, and (iv) while driving in a work zone area increases the risk of property damage crashes, the impact is higher for male drivers compared to female drivers.

*Keywords:* Gender Gap, Crash Injury Severity, Single vehicle crash, Virginia

**INTRODUCTION**

Roadway crashes and injuries inflicted by them have certain demographic trends that need addressing in case of policy development and driver educational campaigns. A distinct trend is visible among male and female drivers in case of injury severity risk (1, 2). These differences mainly stem from differences in driving behavior, exposure, type of vehicle, vehicle condition, driver age, intoxicated driving, and driving conditions. Although there is a significant disparity between the proportion of male and female drivers throughout the world, the impact of crashes is felt through all demographics. According to the World Health Organization (WHO), about 1.25 million people die every year worldwide and 20 to 50 million suffer from injuries due to road traffic related crashes (3). Road traffic injuries are the leading cause of death among people aged 15–44 years and account for 48% of global traffic deaths. Research carried out in 2010 suggests that road traffic crashes cost countries approximately 3% of their Gross Domestic Product which rises to 5% in some low and middle-income countries (3). With this trend, road traffic crashes are predicted to rise to become the 7[th] leading overall causes of human death by 2030 (3). Countermeasures have been developed by identifying important roadway, vehicle characteristics and environmental factors related to crash severity. An important factor that is worth exploring is the difference in significance and magnitude of contributing factors to crash severity in different gender. In the case of the USA, the proportion of female drivers have risen from almost 43% in 1970 to over 50% in 2011. As male and female drivers have observable differences in driving behavior, crash severity analysis considering these differences in population groups provides an opportunity for better understanding of the factors causing gender-gap in injury severity (4-6).

It is well established in the literature that there exists a considerable difference in driving behavior between male and female drivers (5, 7-11). The notion that male drivers take more risk on the road and commit more driving infractions has consensus in the body of work and supported by the fact that male drivers received more traffic citations on the road. Female drivers, on the other hand, have been found to be more cautious in their driving behavior (4, 9). Even with this difference in behavior, a few researchers have concluded that the female drivers are more prone to harmful injuries and fatalities than the male driver (12). Choice of vehicle, roadway geometry, physiology, and driving environment can be attributed to this difference in injury risk. By conducting a comprehensive multivariate analysis, the relative impacts of different attributes on crash injury severity can be observed. Specifically, this paper aims to answer the following questions:

- How factors like the socio-economic, vehicle, temporal, road, and environment (e.g., weather condition and road type) effect crash severity differently among genders?
- How driving behavior and other actions (e.g., intoxicated driving, vehicle maintenance) influence female and male crash severity?
- What factors are the common contributor to increasing/decreasing crash severity among genders, and how do these factors signify the variations among male and female drivers?

Assessing the relative impacts of different factors by means of population segmentation by gender would help transportation agencies in developing targeted policies for safety improvement and driver educational programs. In this paper, police-reported motor vehicle crash data from the state of Virginia collected during the year of 2011-2015 is used. The geo-located crash data along with roadway centerline data is used to develop a comprehensive dataset that includes driver, roadway, environment, vehicle-related, location, collision type, traffic, and driver action related information. The analysis is conducted only on single vehicle crashes. In case of multi-vehicle crashes involving both male and female drivers, many complex interaction variables come into play as The dynamics involved in the multi-vehicle crash can be significantly different than those in the single-vehicle accident case (13). Using all the driver records in multiple-vehicle crashes at different data points cause co-linearity in the input data. Due to these factors, only single vehicle crashes are used in this study. In the Commonwealth of Virginia, the total number of single-vehicle crashes for the reported years were 158,418 with male and female drivers being attributed to 61.4% and 38.6% of the crashes respectively.

The remainder of the study is structured as follows. First, a brief literature review on the methodologies used in crash severity prediction and the impact of different factors on gender-based severity model is provided. Second, a short description of the datasets in question and descriptive statistics of the data is provided. Third, the models developed for this study is discussed in depth along with model structure and the goodness of fit measure appropriateness of the model. Subsequently, the analysis results and some key observations are discussed in detail. Finally, the main takeaway points of this research are discussed along with possible future direction.

## BACKGROUND

One of the most effective way of measuring the relative impact of different factors on injury severity is to model it as the dependent variable with respect to the independent influencing factors (14-17). Crash severity has been modeled in multiple ways in the literature. Most prevalent among them is using parametric models of different forms by using severity as binary, discrete, ordered, and unordered variables. The severity level varies from a binary outcome up to a discrete outcome involving as many as five classes. A wide range of analytical tools have been applied to assess the impact of different factors on both aggregated and disaggregated crash data. Along with the methodological developments, types and quantity of data used in crash severity modeling have increased as more data has become available to researchers. In this section, the methodological alternatives suited for this study and important factors pertains to gender-based severity models are discussed.

One of the most important factors that have been researched in detail, is whether to model crash severities as an ordered discrete outcome or an unordered discrete outcome. The ordered models take into consideration the ordinal nature of injury severity. This order of severity varies from no-injury to possible/severe injury to fatality. Many variations of ordered models have been used in the literature. However, the ordered model suffers from erroneous predictions in case of underreporting of crash severity data and impose a restriction on variables in terms of their influence on the outcome probabilities (18). The unordered discrete outcome model overcomes the restriction put on variables by the ordered model but suffers from the correlation of unobserved effects with the different severity levels (18). The discrete nature of crash data also opens the horizon for developing different non-parametric data analytic models along with parametric ones.

Irrespective of the methodology used, the attributes used in severity prediction modeling is wide-ranging. For the purpose of this study, the focus is placed on researches that have developed a gender-based severity model or use gender as an independent variable. For example, Mercier *et al.* (7, 8) in their research were trying to address whether age or gender or both influenced severity of injuries suffered in head-on and broadside/angle crashes on rural highways respectively. Age and gender were significant variables in these studies with the severity impact being slightly higher for females than males. Use of lap and shoulder restraints seemed more beneficial for men than for women whereas deployed airbags seemed more beneficial for women than for men. Kim *et al.* (19) used age, gender, driver intoxication, driving behavior, and crash type as a predictor of crash severity. In a more comprehensive study, Islam and Mannering (20) used several drivers, vehicle, traffic control, collision type, environment, and roadway information as a predictor of a severity model (20). Although different sets of the variable are used for different severity levels, this study provides a basis for finding important factors for gender-based severity model. In another study, Abigail and Mannering (21) used similar variables as Islam and Mannering (20) and developed several models for different gender and age groups. One important finding of their study was that a different set of variables were found to be significant as a predictor involving age and gender groups. Shen et al. (22) developed a model relating vehicle, driver, environmental, and temporal variables to estimate the health care cost of the crash. Taking health care cost as a surrogate measure of severity with the low cost associated with lower severity crash and vice versa, the factors affecting crash severity can be inferred. The finding of this study is similar to the previous studies as different sets of variables were found to be significant as crash severity predictors (22). Kweon and Kockelman (23) conducted a comparative analysis of crash severity between male and female drivers using collision and vehicle type and found female drivers are more at risk than male drivers; this result is consistent with the study by Abdel-Aty and Abdelwahab (24).

In another comprehensive study, Obeng (1) used several drivers, vehicle, traffic control, collision type, land use pattern, traffic volume, environment, and roadway information to develop an ordered logit model for crash severity prediction. Although this model is specific to intersection crashes, it gives a good estimate of which variables have an impact on cash severity for male and female drivers. One of the more interesting findings of this study is that even though some variables have a similar impact on male and female drivers, their marginal effect varies (1). In another study which is most relevant to the study by, Kim et al. (2), developed a mixed logit model to predict the severity of single-vehicle crashes in California. The authors used different gender and age groups as indicator variables along with crash, vehicle, roadway, environmental, and temporal variables as predictors. Although the study by Kim et al. (2) didn't develop gender-based models, it identifies the variables causing the gender gap in injury severity. Amarasingha and Dissanayake (25) conducted gender-based studies using similar data sets on young drivers in the state of Kansas. Similar to the previous studies, different sets of the variable were found to be significantly related to both male and female drivers.

It is clear from the above studies that injury severity risk has different dynamics in different gender. When separate models are used for severity prediction for different genders, various factors have conflicting and often opposite effect with changing significance level. Variables used in different studies and the trend of those variables on severity risk is summarized in Table 1. In summary, the selection of a methodology to develop a crash severity model is only an initial step for identifying important factors influencing crash severity. Also taking into consideration different age groups, the problem becomes more complicated as trends and level of significance of various factors tend to change from one age and gender group to another. Variable co-linearity and variation in population groups must be taken into careful consideration for modeling purposes. In this paper, these aspects of gender-based modeling are taken into consideration. In this study a two-level nested model has been applied over multinomial logit model structure.

## DATA OVERVIEW

The dataset which is used in this study is the Virginia motor vehicle crash data from 2011 – 2015. This dataset has been acquired through the Traffic Records Electronic Data System (TREDS). Each police reported crash is located by the Center for Geospatial Information Technology at Virginia Tech using information provided by the law enforcement officer on the FR300P crash report (26). This data is a vital component in supporting Virginia's efforts in reducing fatalities, injuries, most importantly crashes and related costs. The data provided by TREDS provides useful and detailed highway safety information for analysis and decision making. There were a total of 611,473 crashes reported by the TREDS system over a period of 5 years period used for this study. Out of these crashes, two-vehicle crashes were the most predominant one, accounting for 60.52% of the total crashes. Single vehicle crashes are the second most frequent accounting for 29.11%. Three, four, and five vehicle crash were 8.43%, 1.55%, and 0.38%, respectively. As stated before, only single vehicle crashes have been used in this study. Traffic and roadway data is collected from roadway centerline data developed by the Virginia Base Mapping Program (VBMP).

### Data Pre-processing

The TREDS crash data consist of four different relational datasets with different attributes. The driver data consists of driver related variables involved in the crashes, the injury data consists of the gender and injury types of the people involved in the crash, and the vehicle data consists of the attributes associated with the vehicle involved in a crash. Finally, the indicator dataset consists of all other variables that can be associated with a crash. One of the most important aspects of the crash data is that it has geographical information which is used to locate the crashes to the exact location on the road. For this purpose, a second data set which provides the road centerline information is used. This dataset was developed as part of the Virginia Base Mapping Program (VBMP). The VBMP roadway centerline data consists of road geometry, roadway type, traffic, speed limit, and locational (rural/urban designation) information for all the public roads in the state of Virginia. Some of this information is attached to the geo-located crash data by means of map matching. Traffic data in form of AADT is attached to the crash data of the same year. This process enriches the available crash information with information about the location and roadway condition of its occurrence.

Initially, the driver data was merged with the vehicle data using a common unique identifier; vehicle driver ID. This combination is further joined with the injury and indicator datasets using another unique identifier. Secondly, this new dataset was joined with the road centerline data using the 'spatial join function' in ArcGIS. Few important attributes obtained from this data enrichment method are traffic volume, lane count, speed limit, road, and roadway surface type.

### Data Exploration

A total of 178,012 single-vehicle crashes occurred in the state of Virginia between the years 2011-2015. Out of these 2,097 were fatal crashes, 63,001 were injury crash and the rest are property damage crashes only. This number includes all motorcycle, pedestrian, moped, bicycle, and train-related crashes. After filtering out all non-motor vehicle and pedestrian crash the total number of crashes dropped to 158,418. The number of crashes that are in each of these five categories by gender is illustrated in Figure 1.

**Table 1:** Factors Used in Gender Based Prediction of Crash Severity and their Relative Impact in

| Variables | Male | | Female | |
|---|---|---|---|---|
| | Increase | Decrease | Increase | Decrease |
| Driver Asleep | (20), (1) | (13) | (20), (1) | (13) |
| Driver Impaired | (1), (10) | (13) | (20), (1) | |
| Driver Ill | (20), (1), (2) | | (20), (1) | |
| Medical Condition | (1) | | (1) | |
| Number of Vehicles in Collision | (1) | | (1) | |
| Head on Collision | (1) | (19) | (19), (1) | (22) |
| Hit Rear of Slow Vehicle | | (1) | | (1) |
| Hit Left Turning Vehicle in the Same Direction | (1) | (13) | | (1) |
| Hit Left Turning Vehicle in Different Direction | | (1) | | (1) |
| Side Swipe | | (1) | | (1) |
| Struck a Pole/Tree | (20) | | (20) | |
| Rear-end crashes | (11) | (1) | | (1) |
| Backing Up | | (1) | | (1) |
| Rollover/Overturned | (20), (19) | | (20), (23) | (19) |
| Passenger Car | (1) | | (23), (1) | |
| SUV | (1) | | (1), (23) | |
| Van | (1) | | (1) | |
| One or More Passengers | (20), (1) | | (13), (1) | |
| Inattention | (13) | | | (13) |
| Old Vehicle (6-10 Years) | | | (20), (13) | |
| Defective Tires Vehicles | | | (13) | |
| Intersection Related | (20) | | (20), (27) | (13) |
| Snowy or Icy Road | | (13), (21) | | |
| Foggy or Snowy Weather | (20) | | (20), (13), (21) | |
| Pickup | (1) | | (1) | |
| Belt Used | | (1) | | (1) |
| Belt Not Used | (20) | (13) | (20) | (13) |
| Airbag Equipped | | (1) | | (1) |
| Airbag Deployed | (1) | | (1), (28) | |
| Wet Road Condition | (13) | (1), (21) | (13), (21) | (1) |
| Driver Age | (20), (2), (19),(6) | | (5) | (9),(6) |
| Driver Trapped/Ejected | (20) | | (20) | |
| Darkness Without Streetlights | (2) | | | |
| Speeding | (20), (2) | (13) | | (13) |

In total, there are 61 different attributes in the combined dataset involving vehicle crashes, including continuous, binary, and categorical. In order to bring consistency in the data, the continuous variables were normalized using a min-max normalization technique where the minimum value of the continuous variable in the dataset is set to zero and the maximum value of the dataset is set to 1. The reason behind this transformation is to have independent variables used for the modeling purpose on a neutral scale. Since most of the variables used for the purpose of this study are discrete variables that are mainly dichotomous, it is imperative to scale continues variables to the same distribution range. It should be noted that other forms of the driver age variables (i.e., binary and categorical age variable for different groups) were tested in the modeling process, but the best results were obtained by using the normalized transformation of the age variable. All other values reside between 0, and 1. All categorical variables were transformed into binary variables representing a single category value as 1 and rest as 0. However, due to the very low frequency, some of the category variables were not transformed into binary variables to restrict a number of variables to a workable amount. After the attribute transformation, a total of 113 variables were prepared as the model predictor. Descriptive statistics of the variables used for the modeling purpose is provided in Table 2.

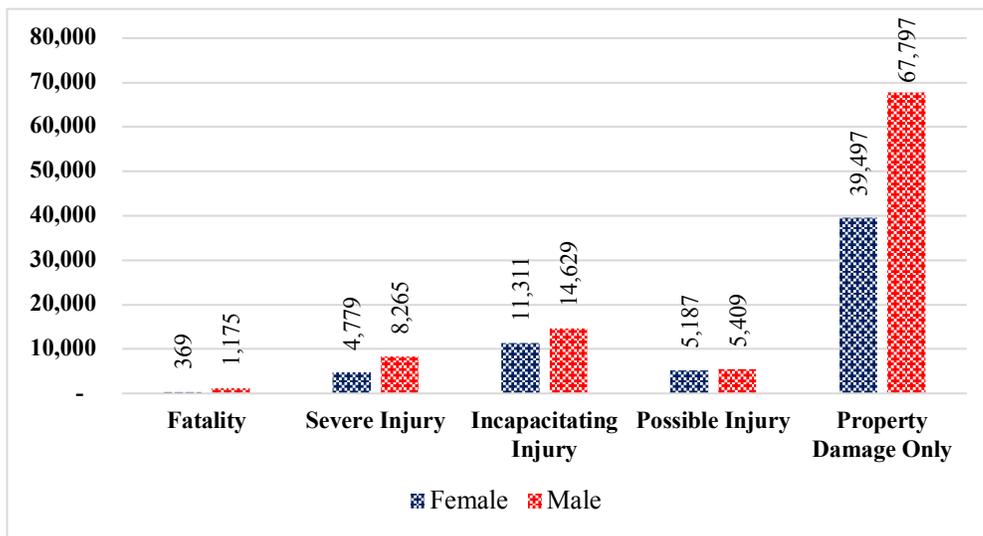

**Figure 1:** Distribution of Severity in Single Vehicle Crashes in Virginia

## MODEL DEVELOPMENT

As discussed above, a nested logit model is selected as the analytical method for this study. As the objective of the study is to find the significant factors affecting gender gap in crash severity risk, the nested structure has an advantage over the multinomial logit model. In case of crash severity estimation, multinomial logit model suffers from the correlation of unobserved effects between the different injury levels. This kind of correlation causes the violation of the IIA property of the model. The IIA property states that selection between alternatives $X$ and $Y$ depend only on the individual preferences between $X$ and $Y$. When the IIA property is violated then the nested logit model is a generalization of the multinomial logit model. In comparison with a sequential logit model, the nested structure allows for correlation of error terms among different severity levels (18). This correlation is measured by means of the Inclusive Value (IV) parameter. However, the structure of the nest has an impact on the significance of the different attribute. The formulation of the nested structure and detailed modeling method is discussed in the following sections.

To account for collinearity between variables the results from bivariate analysis has been used. Selection between 2 highly collinear variables is done based on the Akaike Information Criterion (AIC). Among the candidates, the variable that had higher impact on the model goodness-of-fit is chosen. A total of five injury severity types are selected for developing the nested logit model. To examine the factors affecting gender gap in crash severity two distinct two-level nested logit model has been developed. A model each for male and female drivers involved in single-vehicle crash provides the understanding of the gender gap generators and their relative impact on each gender. The nested structure is shown in Figure 2. This structure is selected by using the model selection criterion proposed by Hesner *et al.* (29). The criterions include (1) goodness-of-fit measure, (2) inclusive value, (3) significance and rationality of the estimate (29, 30). The two-level nested structure used in this study had three limbs in the upper level and two of these levels are then further cleaved into two levels each in the lower level. In the top-level severe injury and fatality are grouped into one nest (Class 1), incapacitating injury and possible injury is regrouped into the second nest (Class 2) and the third nest includes property damage crashes only (Class 3). The first two nests then branch out to the two individual severity levels they consist of. In this structure, it is assumed that fatality and severe injury share some unobserved elements specific between themselves and the same assumption hold true for incapacitating injury and possible injury. It is imperative to note that the nested logit model was developed for male population first and then the same significant parameters were used for female counterparts. The opposite track (i.e. developed the model for female population first) was also tested as well, but the former infers additional significant variables at the set significance level. The probability of one severity outcome based on the two-level NL model is presented in Eq. [1]:

$$P_i = P_n * P_{i|n} = \frac{e^{\frac{1}{\mu}\tau_n}}{\sum_{n' \in N} e^{\frac{1}{\mu}\tau_{n'}}} * \frac{e^{\tau V_i}}{\sum_{i' \in n} e^{\tau V_{i'}}} \qquad [1]$$

Where, $P_i$ is the probability of crash severity outcome $i$, $P_{i|n}$ is the probability of outcome of the limb $n$, $P_{i|n}$ is the probability of crash severity outcome $i$ given it belongs to the limb $n$, $\mu$ is the inverse logsum parameter IV, and $\tau_n = \ln(\sum_i e^{\mu V_i})$ The term $V_i$ can be written as, $V_i = \beta_i X_i$. Where, $X_i$ are the vectors of measurable attributes and $\beta_i$ are vectors of estimable parameters. For model appropriateness, the parameter estimation of the IV must be between zero and one according to McFadden's generalized extreme value derivation (18). If the parameter value of IV is equal to 0, then there exists a perfect correlation among the severity level in the nest. On the other hand, the IV parameter value of 1 implies there is no correlation between the severity level in the nest. To create a straightforward method of finding variables causing the gender gap the male data set is used to identify all the determinants variables first. As the male dataset has more samples than the female one it is expected that male dataset will capture more variability among the attributes. Then for direct comparison, only significant variables in the male model are used to develop the female model.

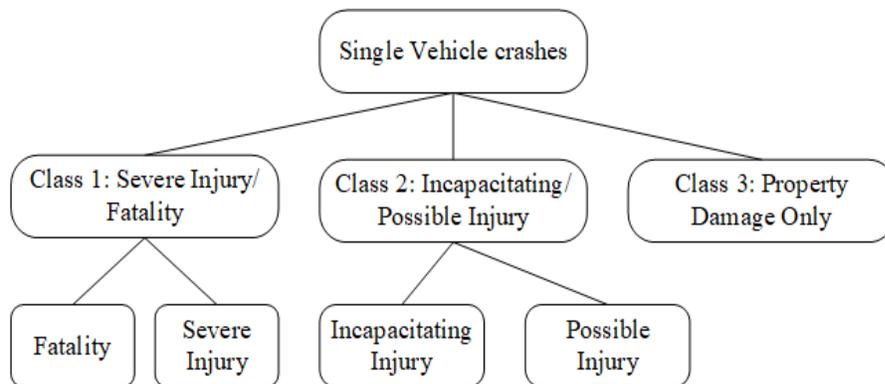

**Figure 2:** Nested Structure

**Table 2:** Descriptive Statistics of Important Attributes

| Variables | Description | Male | | Female | |
|---|---|---|---|---|---|
| | | Avg. | Std. Dev. | Avg. | Std. Dev. |
| Crash severity | 1: Property Damage Only, 2: Incapacitating & Possible Injury, 3: Severe Injury & Fatality | 1.400 | 0.659 | 1.438 | 0.644 |
| Commercial vehicle | 1: Yes, 0: No | 0.041 | 0.199 | 0.003 | 0.054 |
| Speeding | 1: Yes. 0: No | 0.212 | 0.409 | 0.156 | 0.363 |
| Intoxicated driving | 1: Yes, 0: No | 0.160 | 0.367 | 0.081 | 0.273 |
| Crash at intersection | 1: Yes, 0: No | 0.126 | 0.332 | 0.110 | 0.313 |
| Weather: adverse | 1: Fog, Mist, Rain, Snow, Sleet/Hail, Smoke/Dust, Severe Cross Wind and Blowing Sand, soil, dirt or snow, 0: Normal | 0.257 | 0.437 | 0.264 | 0.441 |
| Urban area | 1: Yes, 0: No | 0.523 | 0.499 | 0.505 | 0.500 |
| Driver used drug | 1: Yes, 0: No | 0.021 | 0.145 | 0.019 | 0.135 |
| AADT per lane | Continues variable (Normalized) | 0.043 | 0.071 | 0.042 | 0.068 |
| Age | Continues variable (Normalized) | 0.255 | 0.193 | 0.251 | 0.192 |
| Driver action | 1: Distraction, 2: Avoid Animal, 3: Hit & Run 4: Improper/ Unsafe Lane Change, 5: Fail to Maintain Proper Control, 6: Other | 5.250 | 0.989 | 5.274 | 1.011 |
| Control type | 1: No Passing Lane, 2: Lane Marked, 3: Signal, 4: Stop Sign, 5: Other | 2.692 | 1.408 | 2.625 | 1.392 |
| Driver condition | 1: Fatigue, 2: Sleep, 3: illness, 4: Other | 3.792 | 0.695 | 3.858 | 0.581 |
| Most harmful event | 1: Animal, 2: Overturn, 3: Guardrail, 4: Tree, 5: Ditch, 6: Utility Pole, 7: Other | 4.245 | 2.272 | 4.102 | 2.344 |
| Road surface | 1: Snowy, 2: Wet, 3: Icy, 4: Other | 3.443 | 0.924 | 3.433 | 0.927 |
| Road description | 1: Not Divided, 0: Otherwise | 0.570 | 0.495 | 0.580 | 0.494 |
| Type of collision | 1: Animal, 2: Fixed-Object, 3: Head-On, 4: Angle, 5: Other | 2.293 | 1.160 | 2.179 | 1.123 |
| Road type | 1: Interstate, 2: Primary, 3: Secondary, 4: Urban/City street | 2.452 | 1.030 | 2.473 | 1.007 |
| Hour | 1: [0-3), 2: [3-6), 3: [6-9), 4: [9-12), 5: [12-15), 6: [15-18), 7: [18-21), 8: [21-24) | 4.658 | 2.355 | 4.858 | 2.166 |
| Distraction | 1: Fatigue, 2: Cellphone, 3: Eyes not on road, 4: Other | 3.672 | 0.863 | 3.739 | 0.747 |
| Vehicle body type | 1: Truck-Sport Utility vehicle. 2: Van, 3: Truck-Single Unit truck, 4: Truck-Pickup/passenger truck, 5: Other | 3.992 | 1.461 | 3.953 | 1.668 |
| Vehicle condition | 1: Brake Defective, 2: Slick or Worn Tiers | 0.046 | 0.287 | 0.045 | 0.288 |
| Driver maneuver | 1: Right-Turn, 2: Left-Turn, 3: Runoff road right, 4: Runoff road left, 5: Lane changing, 6: Other | 4.306 | 1.431 | 4.399 | 1.439 |
| Pavement type | 1: BIT, 2: JPC, 3: OTH, 4: UNP, 5: Other | 1.802 | 1.258 | 1.782 | 1.235 |
| Road alignment | 1: Level Curve, 2: Grade Curve, 3: Hillcrest Curve, 4: On or off road, 5: Other | 3.863 | 1.661 | 3.901 | 1.644 |
| Day | 1: Mon, 2: Tue, 3: Wed, 4: Thru, 5: Fri, 6: Sat, 7: Sun | 4.134 | 2.017 | 4.014 | 2.005 |

## RESULTS AND DISCUSSION

The gender-specific models are presented in Table 3. Two datasets separated by gender of the driver have been used to develop the models and the attributes with a significance level of 90% or more are reported in the Table 3. Along with the parameter estimates of the inclusive value parameter, McFadden Pseudo Adjusted $R^2$ value is also reported. As reported in the previous section the range of inclusive value parameter should be between 0 and 1 for model appropriateness. The IV parameter value of severity Class 2 as shown in Table 3. is 0.283 and 0.392 for male and female model respectively. As this value is closer to 0 than 1 it indicates a better correlation among the two-severity level in that class. The same observation holds true of Class 1 severity nest for male's model. In the case of the same class in the female model, the IV value is 0.956 which means there is a low correlation between them but it's within the realm of acceptability.

For the male severity model, in case of property damage only (PDO) crashes the positive coefficient of a variable like icy and snowy roadway condition, work-zone, left turn movement, collision with an animal, and crash on ramps increases the probability of PDO with respect to the other severity. Variables with negative coefficients indicate that those attributes are likely to reduce PDO crashes and may cause more severe injury crash. A single-vehicle crash involving the SUV is more likely to cause a severe injury than PDO. This finding is consistent with previous studies (1, 23). The negative coefficient of driver variable indicates that younger drivers are more likely to suffer severe injuries due to a single vehicle crash. Two different time of the day related variables were found to be significant for male drivers but they were insignificant for female drivers. In both cases, the variables decrease the probability of PDO crashes. The significance of time variable between 9 AM and 12 PM can be explained by more risk-taking tendencies of men in peak period than their female counterparts (10, 21). The time variable between 12 AM and 3 AM can possibly be attributed to late night intoxicated or other impaired driving. This is consistent with similar results obtained for the intoxicated driving variable which was significant only for male. Loss of vehicle control and curve level have a similar impact on both male and female in reducing occurrences of PDO while grade curve was significant only for the male driver.

**Table 3:** Gender Specific Model Results

| Severity Level | Variables | Male Model | | Female Model | |
|---|---|---|---|---|---|
| | | Coefficient | t-test | Coefficient | t-test |
| **Property Damage Only** | *Constant* | 12.08*** | 6.13 | 23.34*** | 5.72 |
| | Roadway Surface: Icy | 0.46*** | 11.25 | 0.36*** | 8.12 |
| | Roadway Surface: Snowy | 0.56*** | 11.49 | 0.53*** | 9.41 |
| | Vehicle Body: Truck Sport Utility | -0.18*** | -8.64 | -0.13*** | -5.97 |
| | Work Zone | 0.44*** | 6.89 | 0.19** | 2.33 |
| | Driver Age | -1.17*** | -19.59 | -1.01*** | -12.87 |
| | Vehicle Maneuver: Left Turn | 0.41*** | 7.47 | 0.36*** | 5.53 |
| | Type of Collision: Animal | 2.97*** | 30.94 | 3.15*** | 29.12 |
| | Crash Occurred Between 12 and 3 am | -0.07** | -2.40 | - | 0.76 |
| | Crash Occurred Between 9pm and 12 am | -0.06** | -2.55 | - | 1.32 |
| | Intoxicated Driving | -0.23*** | -10.06 | - | 1.09 |
| | Driver Action: Fail to Maintain Proper Control | -0.42*** | -17.01 | -0.44*** | -13.17 |
| | Road Alignment: Grade Curve | -1.00*** | -3.28 | - | -1.49 |
| | Road Alignment: Level Curve | -0.93*** | -3.34 | -2.18*** | -2.65 |
| | Road Alignment: On or Off Ramp | 0.51*** | 8.66 | 0.48*** | 6.62 |
| **Possible Injury** | *Constant* | 1.21** | 2.24 | 3.92*** | 3.41 |
| | Speeding Involved | -0.12*** | -4.92 | -0.09*** | -3.39 |
| | Driver Action: Fail to Maintain Proper Control | -0.06*** | -5.86 | -0.06*** | -5.03 |
| | Vehicle Condition: Brakes Defective | 0.05** | 2.38 | - | 1.43 |
| | Vehicle Condition: Worn or Slick Tiers | 0.03*** | 3.31 | - | 1.16 |
| | Crash Occurred at Intersection | 0.31*** | 11.05 | 0.26*** | 8.53 |
| | Driver Age | -0.16** | -2.55 | - | -0.06 |
| | Type of Collision: Animal | 0.77*** | 12.38 | 0.69*** | 10.97 |
| | Crash Occurred Between 12 and 3 am | -0.14*** | -4.79 | -0.07* | -1.91 |
| | Driver Action Improper Unsafe Lane Change | -0.28** | -2.58 | -0.24*** | -2.62 |

| | Value | t | Value | t |
|---|---|---|---|---|
| Weather: Adverse | 0.11*** | 4.32 | 0.08*** | 3.00 |
| Road Alignment: Grade Curve | -0.35*** | -4.22 | -0.37* | -1.85 |
| Road Alignment: Level Curve | -0.32*** | -4.12 | -0.49** | -2.43 |

**Incapacitating Injury**

| | Value | t | Value | t |
|---|---|---|---|---|
| *Constant* | 2.21*** | 4.09 | 4.82*** | 4.19 |
| Speeding Involved | 0.10*** | 9.09 | 0.09*** | 6.79 |
| Crash Occurred at Intersection | -0.17*** | -10.13 | -0.18*** | -8.53 |
| Driver Age | -0.14*** | -4.50 | -0.21*** | -5.55 |
| Driver Action: Fail to Maintain Proper Control | -0.06*** | -5.86 | -0.06*** | -5.03 |
| Driver Vision: Rain Snow on Windshield | -0.05*** | -2.89 | - | -1.21 |
| Vehicle Condition: Brakes Defective | 0.05** | 2.38 | - | 1.43 |
| Vehicle Condition: Worn or Slick Tiers | 0.03*** | 3.31 | - | 1.16 |
| Weather: Adverse | -0.12*** | -8.80 | -0.11*** | -7.57 |
| Road Alignment: Grade Curve | -0.18** | -2.32 | - | -1.32 |
| Road Alignment: Level Curve | -0.18** | -2.52 | -0.49** | -2.46 |

**Severe Injury**

| | Value | t | Value | t |
|---|---|---|---|---|
| *Constant* | 2.61*** | 23.52 | 2.89*** | 18.72 |
| Speeding Involved | 0.15*** | 4.11 | 0.06*** | 3.98 |
| Intoxicated Driving | 0.09** | 2.54 | - | 0.68 |
| Roadway Description: Not Divided | 0.07*** | 4.08 | 0.04*** | 4.24 |
| Type of Collision: Angle | -0.17*** | -3.65 | -0.07*** | -3.24 |
| Speed Limit | 0.002** | 2.60 | 0.001*** | 3.55 |
| Road Alignment: Grade Curve | -0.29*** | -3.64 | - | -1.48 |
| Road Alignment: Level Curve | -0.26*** | -3.64 | -0.33*** | -3.05 |

**Fatality**

| | Value | t | Value | t |
|---|---|---|---|---|
| Roadway Description: Not Divided | 0.07*** | 4.08 | 0.04*** | 4.24 |
| Type of Collision: Angle | -0.17*** | -3.65 | -0.07*** | -3.24 |
| Weather: Adverse | -0.19*** | -4.54 | -0.09*** | -5.03 |
| Speeding Involved | 0.27*** | 4.04 | 0.14* | 1.66 |
| Intoxicated Driving | 0.30*** | 9.43 | 0.37*** | 7.23 |
| Crash Occurred at Intersection | -0.62*** | -4.71 | -1.17*** | -3.75 |
| Vehicle Maneuver: Left Turn | -1.17** | -2.56 | - | 0.00 |
| Speed Limit | 0.01*** | 5.19 | - | 1.40 |

| **Inclusive value parameters** | | | | |
|---|---|---|---|---|
| Class 1 Crash Severity | 0.365*** | 4.65 | 0.956*** | 5.26 |
| Class 2 Crash Severity | 0.283*** | 8.47 | 0.392*** | 7.38 |
| Class 3 Crash Severity | 1 (Fixed) | | 1 (Fixed) | |
| McFadden Pseudo Adjusted $R^2$ | | 0.313 | | 0.284 |
| Sample size | | 97,275 | | 61,143 |

*Note: ***, **, * beside estimated parameter value means significance level at 1 percent, 5 percent, 10 percent consequently.*

In the case of possible injury outcomes, vehicle conditions like defective brake and worn/slick tires are more likely to increase the risk of it happening along with other variables like intersection crash, collision with animals, and adverse weather condition. The findings of the impact of vehicle condition variables are consistent with Ulfarsson *et al*. (13) in case of female drivers but disagree with the male population findings where Ulfarsson *et al.* found no significant relationship. However, two-vehicle crash related variables are not significant for female drivers. Loss of vehicle control, driver age, crash occurrences between 12 and 3 AM, vertical grade curve, and the level curve has a similar negative impact on a possible injury like that on PDO crashes. One notable exception is the driver age which was not found significant in the female model. Speeding and unsafe lane changing variables were found to be decreasing the probability of possible injury crashes and thereby increasing the chance of higher severity crashes.

Incapacitating injury which is in the same nest of severity as possible injury has some similar significant variable in both cases. Speeding, defective brake and tire condition are found be increasing the probability of incapacitating injury. Similar to the two previous severity, vehicle condition related variable is only significant for male driver. Intersection crashes, driver age, loss of vehicle control, adverse weather, vertical grade, and horizontal curve have a similar negative impact on incapacitating injury as possible injury. Vertical grade is not significant in the case of female drivers. Vision-impaired by rain or snow was found significant in male drivers only and it's more likely to reduce incapacitating injury and increase severity risk. Speeding is of particular interest in Class 2 of the severities as they have opposite sign for the estimated parameter in each branch of the class. This indicated presence of speeding is shifting possible injury towards to higher severity of incapacitating injury. This finding agrees with prior research (2, 20). On the other hand, crash at the intersection has the opposite effect in the Class 2 severity level. Intersection crashes are more likely to cause possible injury than incapacitating injury. These findings agree with the findings of Ulfarsson et al. (13) for female drivers but disagree with Islam and Mannering (20) in the case of both genders.

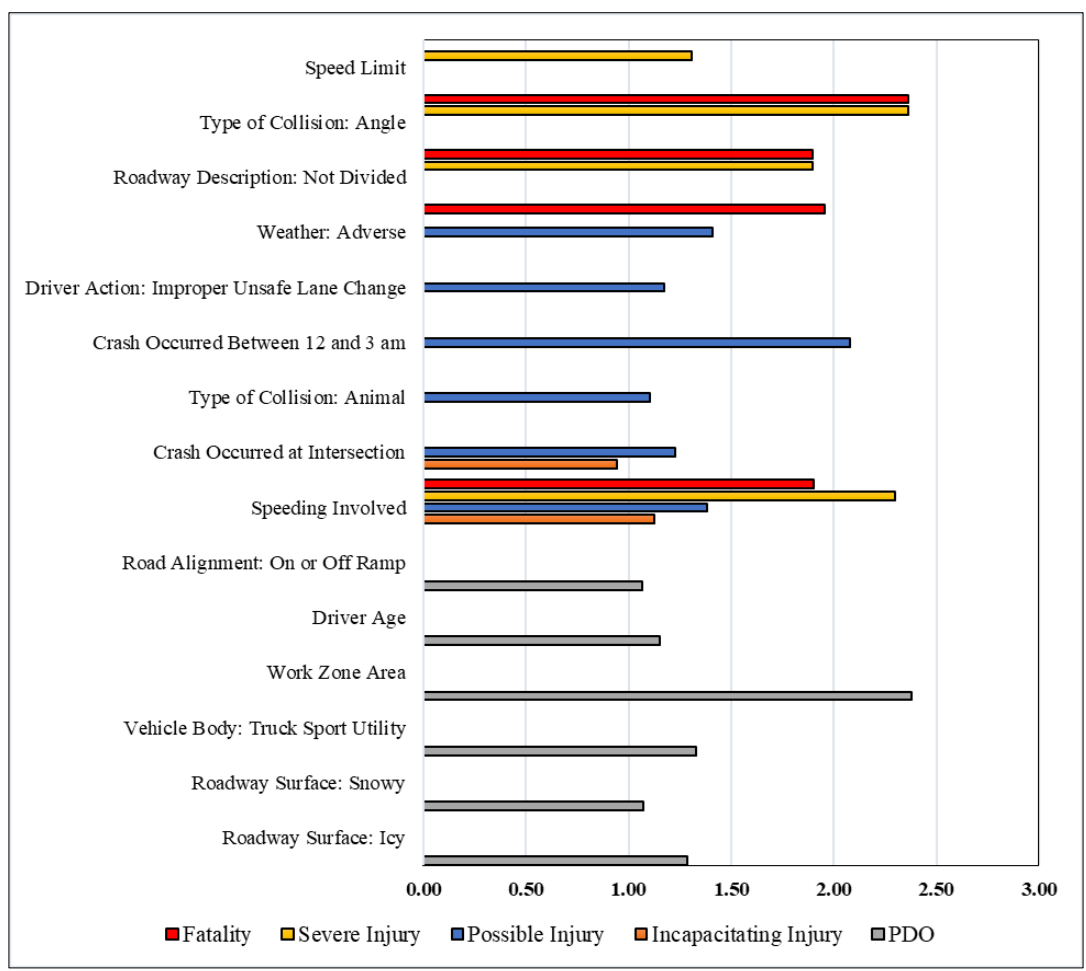

**Figure 3** Ratio of common coefficients for male and female (male coefficient is higher)

In case of crash severity type of Severe Injuries, the number of significant variables is less than any of the lower severity level. Speeding, intoxicated driving, and crash on an undivided roadway are likely to increase the probability of severe injury. Similar to three lower severities, intoxicated driving is only significant for male drivers. Single vehicle crashes on roads with the higher speed limit is significant in both gender and increases the chance of severe crashes. Angled collision, vertical grade, and roadway curve reduce the likelihood of severe injury. Vertical grade is only significant for female driver only.

The likelihood of highest severity level, fatality, is increased by crash occurrences on undivided highway, speeding, intoxicated driving, and increased speed limit. Only in fatal crashes, intoxicated driving is a significant variable in the case of female drivers. Angled collision, adverse weather condition, intersection crash and left turn movement are likely to reduce the occurrences of fatal crashes. In the case of severity group Class 1, the factors that are significant in both branches are intoxicated driving, angled collision type and the speed limit on the roadway. In all cases, the sign of the parameter estimate is the same for both severity levels. This indicates that the impact of the likelihood of both severity levels is the same. Drunk or impaired driving has been found to be increasing the likelihood of both severity level for males and only increases the chance of fatality for female. This finding is consistent with Obeng (1) for both genders and with Islam and Mannering (20) for female drivers only. In contrary, this trend is in contrast with the findings of Ulfarsson *et al.* (13). The impact of speeding is similar to the finding of Islam and Mannering (20) and Kim et al. (2) but doesn't conform with the findings of Ulfarsson *et al.* (13). The significance of intersection related crashes in class 1 is similar to that found in class 2. To provide an in-depth discussion over the gender gap in crash severity, the ratio of coefficients for similar variables in the male and female models are calculated and presented in Figure 3 and Figure 4. Figure 3 depicts all the variables that have a higher impact on male crash severity than female. Variables like the higher speed limit, angled crash, crashes on undivided highway adverse weather condition and speeding has a higher impact on male severity is all severity type. In the case of female drivers, only the roadway curve has a much higher impact when compared to the male counterpart.

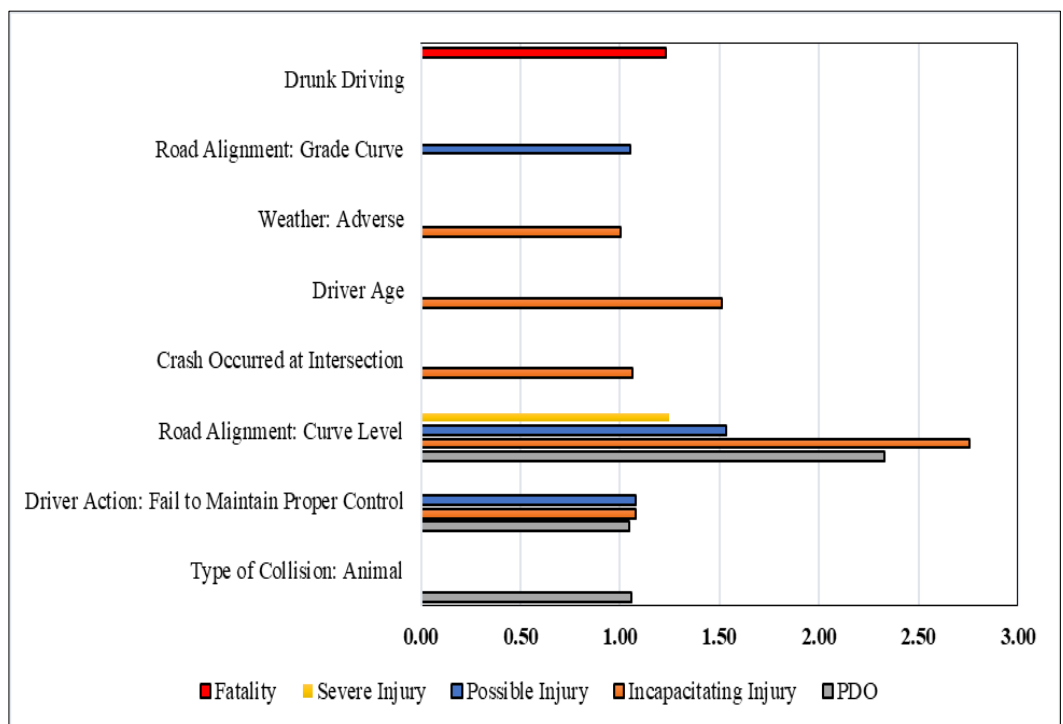

**Figure 4:** Ratio of common coefficients for male and female (Female coefficient is higher)

One important finding in this analysis is the impact of intoxicated driving on both genders. Although, it is only significant for both gender in case of a fatal crash it has more risk on the female driver than male. The actual rate of intoxicated driving related crashes for a male is almost twice than that of the female. That may be an explanation of the significance of intoxicated driving in all severity level for male and not in the female.

## CONCLUSIONS AND RECOMMENDATIONS

This study explores the heterogeneity of crash injury severity in different gender in case of single-vehicle crash occurred in the Commonwealth of Virginia from 2011 to 2015. The main hypothesis was that different driver, environmental, roadway, and vehicle factors impact female and male differently at different severity level. To accomplish this objective, two independent data sources are merged to enrich information related to the crash. These two datasets are the TREDS dataset and VBMP roadway centerline dataset. The TREDS dataset includes geolocated crash data reported by a police officer in FR300P form. The VBMP roadway centerline dataset includes roadway operational, geometrical, and traffic-related information. Combination of these two datasets provides not only driver and vehicle-related information but also the surrounding environment of the crash occurrences. To address the concern about the scalability of different types of variables some contiguous variables are transformed to a neutral scale using a min-max normalization technique.

A two-level nested logit structure is selected as the method of finding variables causing gender gap in crash severity. This method has the advantage of both multinomial logit and ordered logit model by means of allowing for violation of IIA property and correlation of error term among different severity levels respectively. The two-level structure consists of three nests in the upper level. Two of the nests then cleaved into two branches each. In the upper-level fatality and severe injury are grouped into Class 1, Incapacitating and possible injury are grouped into Class 2, and Class 3 contains property damage crashes only. The inclusive value parameter is used to assess the model appropriateness. Nesting structure is kept the same for both genders for direct comparison of variables causing gender-gap in crash severity.

Single vehicle crashes data for male data was used to find the significant variables which are then used to develop a model for females. Late night or midmorning crashes, vehicle defects such as brake failure and worn or slick tires, and driver vision impaired by rain or snow were found to be the contributory factors in male driver only crashes. Female fatality risk in case of intoxicated driving exceeds male risk in this case. However, in all other severity types, the risk is higher for male drivers involved in intoxicated driving crashes. Speeding contributes to lower fatality risk among female drivers compared to their male counterparts. Crashes in the work zone area increase the risk of property damage crashes. The impact is higher for male drivers compared to female drivers. Single vehicle crashes at the intersection are more likely to result in lower severity crash for both genders. The main contribution of this paper is the use of an enriched dataset obtained by combination of multiple datasets in crash severity analysis and application of nested logit structure to observe gender gap in different severity level. A future avenue that can expand on the findings of this study is to incorporate multiple vehicle crashes, so that the complex interaction between the driver and vehicle attributes involved in the same crash can be estimated.

## ACKNOWLEDGMENT

The authors would like to thank the Virginia Department of Motor Vehicles Highway Safety Office for providing the crash data for the analysis. The contents of this paper reflect the views of the authors, who are responsible for the facts and the accuracy of the data and analysis result presented herein. The contents do not necessarily reflect the official views or policies of the Virginia Department of Motor Vehicles.